%% file: kaonKtDep.tex
\documentclass[final,3p,times,twocolumn]{elsarticle}

\makeatletter
\def\ps@pprintTitle{%
  \let\@oddhead\@empty
  \let\@evenhead\@empty
  \let\@oddfoot\@empty
  \let\@evenfoot\@oddfoot
}
\makeatother

\usepackage{amssymb}
\usepackage{lineno}
\usepackage{hyperref}
\usepackage{xfrac}
\graphicspath{{./pictures/}}
\biboptions{sort&compress}

\journal{Physics Letters B}

\begin{document}

\begin{frontmatter}



\title{The transverse momentum dependence of charged kaon 
Bose-Einstein correlations in the SELEX experiment}


\author{The SELEX Collaboration}
\address{}

\include{selex_authors}

\begin{abstract}
We report on the measurement of the one-dimensional charged kaon
correlation functions using 600~GeV/{\it c} $\Sigma^-$, $\pi^-$ and
540~GeV/{\it c} $p$ beams from the SELEX~(E781) experiment at the Fermilab
Tevatron. $K^{\pm}K^{\pm}$ correlation functions are studied for three transverse
pair momentum, $k_T$, ranges and parameterized by a Gaussian form.
The emission source radii, $R$, and the correlation strength,
$\lambda$, are extracted. The analysis shows a decrease of the source
radii with increasing kaon transverse pair momentum for all beam types.
\end{abstract}


\end{frontmatter}


\section{Introduction}
\label{intro}
In this paper we present results from $K^{\pm}K^{\pm}$ correlation
femtoscopy study in $600$~GeV/{\it c} $\Sigma^- C(Cu)$, $\pi^- C(Cu)$ 
and $540$~GeV/{\it c} $p C(Cu)$ interactions from the SELEX~(E781) 
experiment~\cite{selex} at the Fermilab Tevatron. The correlation femtoscopy
method allows to study spatio-temporal characteristics of
the emission source at the level of $1$~fm = $10^{-15}$~m.
The method is based on the Bose-Einstein enhancement 
of identical boson production at small relative momentum. 
The quantum statistics correlations were first observed as an enhanced
production of the identical pion pairs with small opening angles in
proton-antiproton collisions~\cite{gglp1}. In 1960 this enhancement
was explained by the symmetrization of the two-particle wave function
by G.~Goldhaber, S.~Goldhaber, W.-Y.~Lee, and A.~Pais (GGLP effect)~\cite{gglp2}.
Later, in the 1970s, Kopylov and Podgoretsky suggested studying
the interference effect in terms of the correlation function. 
They proposed the mixing technique to construct the uncorrelated
reference sample, and clarified the role of the space-time
characteristics of particle production~\cite{kp1, kp2, kopylov}.
Subsequently, two-particle correlations at small relative momentum were
systematically studied for lepton-lepton~\cite{opal1}, lepton-hadron~\cite{zeus1},
hadron-hadron~\cite{aamodt1}, and heavy-ion~\cite{na44_1, star4}
collisions. It was found that the system created in heavy-ion
collisions undergoes the collective expansion and may be described by
relativistic fluid dynamics~\cite{star_hydro, phenix_hydro,
phobos_hydro, brahms_hydro}. By using the width of the
quantum statistical enhancement, one can measure the radii $R$ of the
emitting source. The decrease of the extracted radii with increasing
transverse pair momentum may be interpreted as the decrease of the
``homogeneity lengths''\cite{homogeneity} due to collective transverse
flow.

A comparison of femtoscopic measurements in lepton and 
hadron-induced~\cite{alexander4, kittel} collisions with heavy-ion
collisions shows similar systematics~\cite{chajecki2, aggarwal}.
These studies usually performed for pions. However, measurements of
heavier particles may provide additional information about the size,
orientation and dynamical timescales of the emission region.

\section{Experimental setup and data selection}

The SELEX (E781) detector is a three-stage magnetic spectrometer 
designed for charm hadroproduction study at $x_F > 0.1$ ($x_F=\frac{p_Z}{p_{Zmax}}$).
We report on the analysis of 1 billion events of $600$~GeV/{\it c} $\Sigma^-C(Cu)$, 
$\pi^-C(Cu)$ and $540$~GeV/{\it c} $p~C(Cu)$ interactions recorded
during the 1996--1997 fixed target run. About 2/3, 1/6 and 1/6 of the
data were obtained on $\Sigma^-$, $\pi^-$ and $p$ beams, respectively.

The beam particle was identified as a meson or a baryon by a transition 
radiation detector. Interactions occurred on segmented targets,
which consisted of 2~copper and 3~diamond foils separated by 1.5~cm 
clearance, and had a total thickness of 5\% of an 
interaction length for protons. Particles were tracked in a
set of 20 vertex Silicon Strip Detectors (SSD) arranged in 4 sets 
of planes with a strip pitch of 20--25~$\mu$m, rotated by $45^{\circ}$.
Each of the detectors has a hit detection efficiency greater than 98\%.
Transverse vertex position resolution ($\sigma$) was 4~$\mu$m for the
600~GeV/{\it c} beam tracks. The average longitudinal vertex position
resolution was 270~$\mu$m for primary vertex and 560~$\mu$m for
secondary vertex. The detector geometry covers the forward 150 mrad cone.
The particle momentum was measured by deflection of the track position by two 
magnets M1 and M2 in a system of proportional wire chambers and 
silicon strip detectors. Momentum resolution of a typical 100~GeV/{\it
c} track was $\sigma_p/p \approx 0.5\%$. A Ring Imaging Cherenkov
detector (RICH) performed particle identification in a wide momentum 
range and provided 2~$\sigma$ $K/\pi$ separation up to 
165~GeV/{\it c} and single track ring radius resolution of
1.4\%~\cite{engelfried1}. The kaon identification efficiency was
over 90\% above the kaon threshold ($\approx$43~GeV/{\it c}). The
average number of tracks reaching the RICH was about 5 per event~\cite{engelfried2}.
The layout of the spectrometer is described elsewhere~\cite{selex}.

\begin{figure}[h]
  \includegraphics[width=\columnwidth]{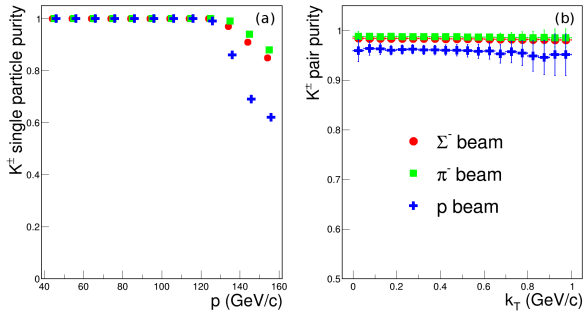}
  \caption{Single $K^{\pm}$ purity as a function of momentum (a) and
    the average transverse pair momentum dependence of the 
    $K^{\pm}$ pair purity for the $\Sigma^-$ (circles), $\pi^-$ (squares) and p
    (crosses) beams (b).}
  \label{figKaonPurity}
\end{figure}

In this analysis we used primary tracks that have vertex silicon track
segment matched with downstream segments measured in the M1 and M2 spectrometers, with
the momentum from 45 to 160~GeV/{\it c}. In order to reduce the contamination
of secondary particles, it was required that the extrapolated track distance
to the primary vertex was less than 15~$\mu$m in the transverse plane.
Only tracks that matched the RICH detector were used in the analysis.
Charged kaons were identified with the likelihood to be a kaon at least 
three times exceeding any other particle hypothesis. 
Fig.~\ref{figKaonPurity}(a) shows the single kaon purity
as a function of the momentum for the $\Sigma^{-}$, $\pi^{-}$ and $p$
beams. It is defined as the fraction of the accepted kaon
tracks that correspond to true kaon particles. The single particle
purity was estimated from the RICH ring radius distributions 
of the data and by studying PYTHIA~\cite{pythia} simulations with
the particle embedding through the detector. The main contamination for
charged kaons in the momentum range $p>120$~GeV/{\it c} comes from pions,
because of the close ring radii in the RICH detector.

The electrons were eliminated from the analysis using the
transition radiation detector (ETRD) that was placed before RICH.
The contamination from other particle species in the studied momentum range
is negligible. Fig.~\ref{figKaonPurity}(b) shows the charged kaon pair
purity as a function of the average transverse momentum of the pair
obtained for the $\Sigma^{-}$, $\pi^{-}$ and $p$ beams.
The $K^{\pm}$ pair purity is calculated as a product of two
single-particle purities using the experimental momentum distributions.

After applying the cuts 4842147, 597101 and 103551 identical charged kaon
pairs were selected for $\Sigma^-$, $\pi^-$ and $p$ beams, respectively.

\section{Correlation femtoscopy}

The two-particle correlation function is defined as the ratio
of the probability to measure two particles with momenta $\vec{p}_1$
and $\vec{p}_2$ to their single particle probabilities:
\begin{equation}
C(\vec{p}_1,\vec{p}_2) = \frac{P(\vec{p}_1,\vec{p}_2)}{P(\vec{p}_1)P(\vec{p}_2)} .
\end{equation}
Experimentally, one studies the correlation function
$C(\vec{q},\vec{K})$ in terms of relative momentum
$\vec{q}=\vec{p}_1-\vec{p}_2$ and average momentum
$\vec{K}=(\vec{p}_1+\vec{p}_2)/2$ of two particles:
\begin{equation}
C(\vec{q},\vec{K}) = \frac { A(\vec{q},\vec{K}) } { B(\vec{q},\vec{K}) } 
\cdot D(\vec{q},\vec{K}) , \label{corrFctnDef}
\end{equation}
where $A(\vec{q},\vec{K})$ is the measured distribution of relative momentum 
within the same event, $B(\vec{q},\vec{K})$ is the reference or background
distribution that is similar to the experimental distribution in all
respects except for the presence of femtoscopic correlations. The
reference sample is usually formed from particles that come from different
events (event mixing technique~\cite{kopylov}). The quantity
$D(\vec{q},\vec{K})$ is a so-called correlation baseline that
describes all non-femtoscopic correlations, such as, for instance, the
correlations caused by the energy and momentum conservation-induced
correlations~\cite{chajecki}. 
In order to eliminate possible biases due to
the construction of the reference samples, the measured correlation
functions were corrected on the simulated distributions by
constructing the double ratio:
\begin{equation}
  \small
  R(Q) = \left( \frac {dN_{K^{\pm}K^{\pm}}/dQ} {dN_{ref}/dQ} \right) /
  \left( \frac {dN_{MC,K^{\pm}K^{\pm}}/dQ} {dN_{MC,ref}/dQ} \right),
  \label{eqDoubleRatio}
\end{equation}
where the subscripts ``$MC$'' and ``$MC,ref$'' correspond to the
simulated data.

By virtue of the limited statistics available for the $\pi^-$ and $p$
beams, only the one-dimensional femtoscopic charged kaon analysis of
correlation functions in terms of invariant relative momentum, 
$Q = \sqrt{ (\vec{p}_1-\vec{p}_2)^2 - (E_1-E_2)^2 }$, was performed.
In order to extract the size of the emission region, $R$, one can use
the Goldhaber parametrization. This assumes that the emitting source
of identical bosons is described by a spherical Gaussian density
function:
\begin{equation}
C(Q) = N \left( 1 + \lambda e^{-R^2Q^2} \right) \cdot D(Q) , \label{gaussian}
\end{equation}
where $N$ is a normalization factor, $\lambda$ describes the
correlation strength, and $D(Q)$ is the baseline distribution.
In the current analysis, the second order polynomial, $D(Q)=1+aQ+bQ^2$,
was used for estimation of the baseline distribution. 
The momentum correlations of particles emitted at nuclear distances
are also influenced by the effect of final-state interaction (FSI),
Coulomb and strong interactions~\cite{lednicky, voloshin,
  pratt_fsi, lednicky2}. For identical kaons, the effect of strong
interactions is negligible~\cite{lednicky4}. The correlation function of
identical bosons should increase at low relative momentum, except for
small values where Coulomb interaction becomes dominant. This may be taken
into account by modifying Eq.~(\ref{gaussian}):
\begin{equation}
  C(Q) = N \left( (1 - \lambda)+\lambda K(Q) \left( 1 + 
  e^{ -R^{2}Q^{2} } \right) \right) \cdot D(Q) , \label{BowSinFit}
\end{equation}
where the factor $K(Q)$ is the squared like-sign kaon pair
Coulomb wave function integrated over a spherical Gaussian 
source~\cite{bowler, sinyukov, star_coulomb}.

\section{Results and discussions}

The results discussed in this Letter were obtained
with the same detector setup, cuts and fitting procedures,
giving an opportunity to compare the properties of
the emission region for different hadron-induced collisions.
The analysis was performed for three average transverse pair
momentum $k_T = \left|\vec{p}_{T1}+\vec{p}_{T2} \right|/2$ ranges: 
($0.00-0.30$), ($0.30-0.55$), ($0.55-1.00$) GeV/{\it c} 
and for the three beam types: $\Sigma^-$, $\pi^-$, $p$.
The event mixing technique was used to construct the
uncorrelated reference sample. Only events with two or more identical
charged kaons, grouped by production target, were used in the event
mixing. Kaons from adjacent events for each target were combined to
provide an uncorrelated experimental background. Due to small
differences in the measured correlation functions, the positive and
negative kaon four-momentum distributions were combined in the
numerator and the denominator before constructing the ratio. A purity
correction was applied to the experimental correlation functions 
according to the expression:
\begin{equation}
  C(Q) = \frac {C_{experimental}(Q)-1} {P(Q)} + 1 ,
  \label{eqPPCorrection}
\end{equation}
where $P(Q)$ is the pair purity.

\begin{figure}[h]
  \includegraphics[width=\columnwidth]{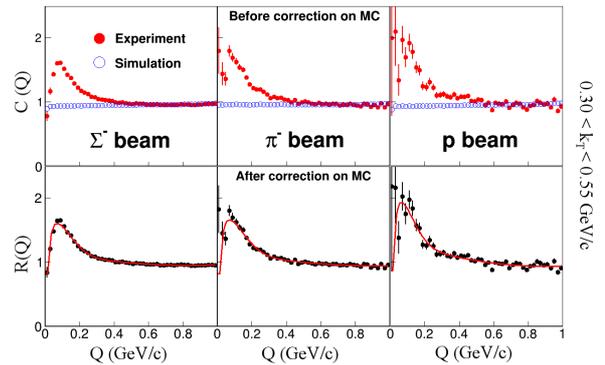}
  \caption{Top row shows the experimental (full circles)
    and PYTHIA-generated (open circles) correlation functions of
    identical kaons obtained from using the event mixed reference
    samples. The bottom row shows the fits to the double ratios
    according to Eq.~(\ref{BowSinFit}). The second order polynomial
    was used for estimation of the non-femtoscopic effects. The
    correlation functions are  measured for $0.30<k_{T}<0.55$~GeV/{\it
    c} range and columns from left to right represent the data
    obtained for $\Sigma^-$, $\pi^-$ and $p$ beams, respectively.
  }
  \label{figCompsAndFits}
\end{figure}

The top row of the Fig.~\ref{figCompsAndFits} shows the experimental
correlation functions (solid circles) after applying the purity
correction measured in the $0.30<k_T<0.55$~GeV/{\it c} region for $\Sigma^-$,
$\pi^-$ and $p$ beams. The correlation functions were normalized such
that $C(Q)=1$ for $0.5<Q<0.7$~GeV/{\it c}, where Bose-Einstein
correlations are absent and the influence of the non-femtoscopic
effects is small. The deviation of the correlation functions from
unity at $Q>0.7$~GeV/{\it c} correspond to the non-femtoscopic
correlations and the well-defined enhancement at $Q<0.4$~GeV/{\it c} is
due to the quantum statistical correlations. The Coulomb repulsion
between like-sign kaons leads to the decrease of the correlation
functions at $Q<0.1$~GeV/{\it c}.

In order to correct for non-femtoscopic effects, the Monte Carlo
event generator PYTHIA-6.4.28~\cite{pythia} with different tunes
(Perugia~0, Perugia~2010 and Perugia~2011~\cite{perugia}) was
used. PYTHIA contains neither Bose-Einstein correlations nor
the final-state interactions. On the other hand, PYTHIA contains
other kinematic effects, for instance, energy and momentum
conservation effects, that could lead to baseline correlations.
The Perugia~2011 tune, which best describes charged-particle
multiplicity, was used as the main minimum-bias MC sample.
The top row of the Fig.~\ref{figCompsAndFits} shows the comparison of
simulated correlation functions (empty circles), where PYTHIA events
were filtered through the analysis cuts, with the experimental
distributions (solid circles) measured for $0.30<k_{T}<0.55$~GeV/{\it c} range. 
The PYTHIA-generated correlation functions were normalized in the same
way as the experimental correlation functions.

It is seen that PYTHIA qualitatively describes the experimental
baseline in the region $Q>0.5$~GeV/{\it c}, where the effect of
femtoscopic correlations is negligible. Since the MC calculation does
not include wave function symmetrization for identical particles, the
femtoscopic peak at low relative four-momentum region,
$Q<0.4$~GeV/{\it c}, is absent. 

The bottom row of the Fig.~\ref{figCompsAndFits} shows double ratios,
where the experimental correlation functions are divided by the
PYTHIA-generated ones, obtained for the $\Sigma^-$, $\pi^-$ and $p$
beams in the $0.30<k_{T}<0.55$~GeV/{\it c} region. The double ratios were
fitted using Eq.~(\ref{BowSinFit}). In the current analysis, the
Coulomb function $K(Q)$ was integrated over a spherical source of
1~fm. Due to imperfections of the simulation in the
$Q>0.7$~GeV/{\it c} region, the non-femtoscopic term $D(Q)=1+aQ+bQ^2$
was used.

\begin{table*}[t]
  \centering
  \caption{Systematic uncertainty (minimal and maximal) values for
    different sources of systematic uncertainty (in percent). The
    values of minimum (maximum) uncertainty can be from different
    average transverse pair momentum range or the beam type, but from a
    specific source.}
  \label{tabSystUncert}

  \small
  \begin{tabular*}{0.6\textwidth}{@{\extracolsep{\fill}}ccc}
    \hline
    The systematic uncertainty source & $\lambda$ (\%) & $R$ (\%) \\
    \hline
    Event/particle selection & 1--7 & 1--9 \\
    PID and purity        & 0--4  & 0--6  \\
    Fit range             & 1--5  & 1--4  \\
    Momentum resolution   & 0--1  & 0--1  \\
    Two-track effects     &  --   &  --   \\
    Non-femtoscopic form  & 0--4  & 1--11 \\
    Coulomb function      &  --   &  --   \\
    Reference sample      & 1--8  & 5--13 \\
    \hline
    \textbf{Total}        & 2--13 & 5--21 \\
    \hline
  \end{tabular*}
\end{table*}

\begin{figure}[h]
  \includegraphics[width=\columnwidth]{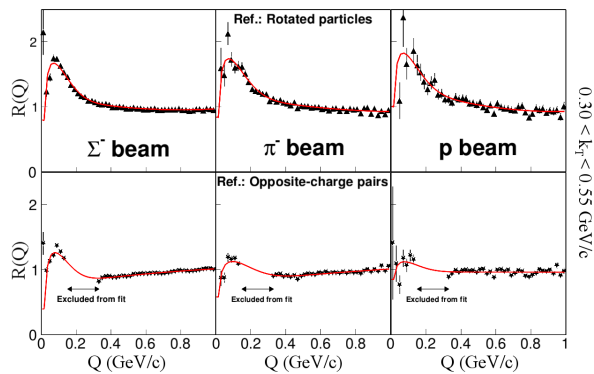}
  \caption{The correlation functions constructed with rotated
    particles (top panel) and opposite-charge pairs (bottom panel) for
    $0.30<k_{T}<0.55$~GeV/{\it c} range. The range
    $0.19<Q<0.35$~GeV/{\it c} on the bottom panel is excluded from
    fits due to the contribution from the $\phi(1020)$ meson
    decay. The fits were performed using Eq.~(\ref{BowSinFit}). The
    columns represent $\Sigma^{-}$ (left), $\pi^{-}$ (middle) and $p$
    (right) beams.
  }
  \label{figDiffRefSamples}
\end{figure}

To estimate the influence of choice of the reference sample,
the different methods of constructing uncorrelated charged particle
distributions were used: {\it opposite-charge pairs} and {\it rotated
particles}, where pairs are constructed after inverting the $x$ and $y$
components of the three-momentum of one of the two particles.
Fig.~\ref{figDiffRefSamples} shows the double ratios obtained from using
rotated particles (top panel) and opposite-charge pairs (bottom
panel) reference samples. The double ratios were fitted using
Eq.~(\ref{BowSinFit}); and the second order polynomial,
$D(Q)=1+aQ+bQ^2$, was used to describe the non-femtoscopic term.
It was found that the extracted femtoscopic parameters, $\lambda$ and
$R$, obtained from using rotated particles reference samples are similar to
those from the event mixed ones.

The reference samples constructed from the opposite-charge kaon pairs
contain peaks coming from strong resonance decays; and are influenced by
the Coulomb attraction at $Q<0.1$~GeV/{\it c}. The magnitudes of the
resonance peaks measured for the real and simulated correlation
functions were found to be different. This can be explained by the
absence of the final-state rescattering of particles in PYTHIA. The
double ratios obtained for unlike-sign kaon pairs were fitted the same
way as the event mixed ones. The range $0.19<Q<0.35$~GeV/{\it c} was excluded
from fits due to the $\phi(1020)$ meson decay and because the influence of the
final-state interactions between opposite-charge kaon pairs was not
taken into account.

The different sources of systematic uncertainties were studied
for each $k_T$ range and beam type. Tab.~\ref{tabSystUncert} shows the
maximal and minimal values of systematic uncertainty that correspond
to specific uncertainty sources. The values of the total uncertainty
are not necessarily equal to the quadratic sum of all the
uncertainties due to the fact that they can come from different beam
types or average transverse pair momentum ranges.

The uncertainty due to the fit range was estimated by varying the upper
limit of the fit from $Q=1$~GeV/{\it c} to $Q=0.6$~GeV/{\it c}. The lowest
value of the upper limit of the fit range corresponds to the end of
the correlation region. Changing the radius of the Coulomb source in
the range from 0.5~fm to 1.5~fm has negligible effect on the extracted
emitting source parameters.

Different baseline shapes~\cite{malinina, cms, akkelin} were used to
estimate the systematic uncertainty due to the baseline determination:
\begin{equation}
  D(Q) = 1 , \label{constTail}
\end{equation}
\begin{equation}
  D(Q) = 1 + aQ , \label{linearTail}
\end{equation}
\begin{equation}
  D(Q) = 1 + e^{-aQ^2}, \label{gausTail}
\end{equation}
\begin{equation}
  D(Q) = \sqrt{ 1 + aQ^2 + bQ^4} . \label{sqrtTail}
\end{equation}

The smearing of single particle momenta was studied by embedding
simulated kaon tracks with known momenta through the detector.
Experimental correlation functions were corrected for momentum
resolution using the expression:
\begin{equation}
  C_{corr}(Q) = \frac {C_{uncorr}(Q)C_{unsmeared}(Q)} {C_{smeared}(Q)} ,
\end{equation}
where $C_{corr}(Q)$ is the corrected correlation function and
$C_{uncorr}(Q)$ represents the measured correlation function.
The $C_{unsmeared}(Q)$ and $C_{smeared}(Q)$ are the correlation functions
without and with taking into account the momentum resolution effect,
respectively. The smearing of the single particle momenta leads to the
smearing of the correlation function in the $Q<0.2$~GeV/{\it c} range.
This decreases the amplitude of the peak and makes it wider. The systematic
uncertainty of this effect does not exceed 1\%.
The two-track reconstruction effects: ``merging'', when two tracks are
reconstructed as one; and ``splitting'', when one track is
reconstructed as two, were studied and found to be negligible.

The main systematic uncertainty arises from using different methods of
constructing the uncorrelated reference sample. The dominant
contribution comes from using opposite-charge kaon pairs.
The contamination of the ``Reference sample'' uncertainty
from using different PYTHIA tunes does not exceed $5\%$.

The systematic errors on $R$ and $\lambda$ for each beam type and
$k_T$ range are taken as the rms spread of the values obtained for
the different sources of systematic uncertainty.

\begin{table*}[t]
  \centering
  \caption{$K^{\pm}K^{\pm}$ source parameters for $\Sigma^-$,
    $\pi^-$ and $p$ beams. Statistical and systematic uncertainties
    are presented.}
  \label{tabCorrVsKt}

  \small
  \begin{tabular*}{0.99\textwidth}{@{\extracolsep{\fill}}cccccccc}
    \hline
    Beam type & $k_T$~(GeV/$c$) & $\chi^2$/$N_{dof}$ & $N$ & $\lambda$
    & $R$~(fm) & $a$ (GeV/$c^{-1}$) & $b$ (GeV/$c^{-2}$)\\ 

    \hline
               & $(0.00-0.30)$ & $126/45$ & $1.23\pm0.01$ & $0.71\pm0.02\pm0.08$ & $1.32\pm0.02\pm0.07$ &
               $-0.59\pm0.02$ & $0.38\pm0.02$ \\
    $\Sigma^-$ & $(0.30-0.55)$ & $85/45$ & $1.18\pm0.01$ & $0.66\pm0.02\pm0.08$ & $1.18\pm0.02\pm0.05$ & $-0.47\pm0.03$ &
    $0.28\pm0.02$ \\
               & $(0.55-1.00)$ & $142/45$ & $1.05\pm0.03$ & $0.66\pm0.04\pm0.10$ & $0.98\pm0.03\pm0.04$ &
               $-0.22\pm0.08$ & $0.13\pm0.05$ \\
    
    \hline 
            & $(0.00-0.30)$ & $62/45$ & $1.19\pm0.03$ & $0.67\pm0.06\pm0.09$ & $1.25\pm0.06\pm0.06$ &
            $-0.49\pm0.07$ & $0.31\pm0.05$ \\
    $\pi^-$ & $(0.30-0.55)$ & $66/45$ & $1.21\pm0.04$ & $0.69\pm0.06\pm0.06$ & $1.13\pm0.06\pm0.06$ & $-0.46\pm0.09$ &
    $0.24\pm0.07$ \\
            & $(0.55-1.00)$ & $58/45$ & $1.34\pm0.07$ & $0.44\pm0.10\pm0.11$ & $1.16\pm0.19\pm0.14$ &
            $-0.71\pm0.09$ & $0.42\pm0.09$ \\

    \hline 
        & $(0.00-0.30)$ & $65/45$ & $1.51\pm0.06$ & $0.98\pm0.17\pm0.13$ & $1.54\pm0.16\pm0.17$ & $-0.97\pm0.10$ &
        $0.62\pm0.08$ \\
    $p$ & $(0.30-0.55)$ & $62/45$ & $1.39\pm0.12$ & $0.80\pm0.15\pm0.13$ & $1.32\pm0.12\pm0.15$ & $-0.72\pm0.13$ &
    $0.40\pm0.11$ \\
        & $(0.55-1.00)$ & $43/44$ & $1.26\pm0.16$ & $0.91\pm0.24\pm0.11$ & $1.13\pm0.17\pm0.11$ & $-0.61\pm0.31$ &
        $0.37\pm0.24$ \\

    \hline
  \end{tabular*}
\end{table*}

\begin{figure}[h]
  \includegraphics[width=\columnwidth]{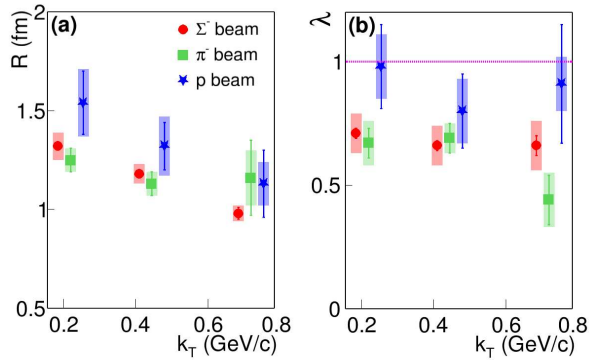}
  \caption{$K^{\pm}K^{\pm}$ source parameters $R$ (a) and $\lambda$ (b)
    measured for $\Sigma^-$ (circles), $\pi^-$ (squares) and $p$
    (stars) beams as a function of transverse pair momentum,
    $k_T$. Statistical (lines) and systematic (boxes) uncertainties
    are shown.}
  \label{figKtDep}
\end{figure}

The results of fits of $R(Q)$ based on the parametrization of
Eq.~(\ref{BowSinFit}) with $D(Q)=1+aQ+bQ^2$ are given in
Table~\ref{tabCorrVsKt}. The extracted source radii and
correlation strength for $\Sigma^-$ (circles), $\pi^-$ (squares) and
$p$ (stars) beams as a function of transverse pair momentum are shown in
Figs.~\ref{figKtDep}(a) and \ref{figKtDep}(b), respectively.
The source radii slightly decrease with increasing $k_T$ for all the
beam types, except the highest average transverse pair momentum
interval for the $\pi^-$ beam. The femtoscopic radii measured for
$\Sigma^-$, $\pi^-$ and $p$ beams are consistent within the
uncertainties. The small difference between measured source parameters
probably arises from different contamination from resonance
decays~\cite{lednicky3}. 

The decrease of the source radii with transverse pair momentum
was previously observed in heavy-ion collisions and interpreted
as a collective hydrodynamic behavior (collective flow)~\cite{lisa, pratt}.
The first direct comparison of correlation femtoscopy in $p+p$ 
and heavy-ion collisions under the same detector conditions, 
reconstruction, analysis and fitting procedures was performed 
by the STAR collaboration~\cite{aggarwal}. It was shown that $p+p$ 
collisions also have the transverse momentum scaling. 
Although the interpretation of these results is still
unclear, the similarities could indicate a connection
between the underlying physics.

The transverse momentum dependence was also observed 
for $\pi\pi$ in $e^+e^-$~\cite{opal3,l3} 
and $pp$ collisions~\cite{aggarwal, cms, aamodt2}. For the 
first time a similar analysis of charged kaon Bose-Einstein 
correlations for more than one transverse pair momentum 
and multiplicity range was recently performed by the ALICE 
collaboration at the LHC in $pp$ collisions at 
$\sqrt{s}=7$~TeV~\cite{malinina}. It was shown that charged kaon
femtoscopic radii decrease with transverse pair momentum
for middle and high multiplicity ranges.

There are several possible processes that may lead to the $k_T$ 
dependencies in the hadronic collisions:

1. The space-momentum dependence of the femtoscopic radii may 
be generated by long-lived resonances~\cite{wiedemann}.
In particular this may play a significant role in
high multiplicity bins, where the bulk collective
flow is predicted~\cite{werner}.

2. Humanic's model~\cite{humanic}, based on space-time geometry of
hadronization and effects of final-state rescattering between hadrons, 
reproduces both multiplicity and transverse mass dependence measured
at the Tevatron~\cite{alexopoulos}.

3. In small systems, the string fragmentation should 
generate momentum and space correlations, such as $k_T$
dependence of the source radii. However, there are almost
no quantitative predictions that may be directly compared
with data except the $\tau$-model in which space-time and momentum
space are strongly correlated~\cite{csorgo}. 
Moreover, the Lund string model is not able to reproduce the 
mass dependence of the radii~\cite{alexander4, bialas1, alexander1, alexander2}.

4. Hydrodynamic bulk collective flow may lead to a $k_T$
dependence that is very similar to that from heavy-ion
collisions.

Taking the aforementioned possibilities, the origin of the transverse pair
momentum dependence of the femtoscopic radii in hadronic collisions 
is still unclear. Further theoretical studies are needed in order
to understand the underlying physics.

\section{Summary}
Charged kaon Bose-Einstein correlations were measured in the SELEX
experiment. One-dimensional charged kaon correlation functions in
terms of the invariant four-momentum difference were constructed for
$\Sigma^-$, $\pi^-$ and $p$ beams and three transverse pair momentum
ranges: ($0.00-0.30$), ($0.30-0.55$) and ($0.55-1.00$)~GeV/{\it c}. 
The source parameters of correlation strength, $\lambda$, and 
source radii, $R$, were extracted for all beam types and for the three  
average transverse pair momentum ranges.
The slight decrease of the femtoscopic radii with pair 
transverse momentum was observed for all three beam types, except for
the highest $k_T$ range of the $\pi^-$ beam. The values of the
emitting source radii obtained for $\Sigma^-$, $\pi^-$ and $p$ beams
are consistent within the uncertainties.

\section*{Acknowledgments}
The authors are indebted to the staff of Fermi National Accelerator Laboratory
and for invaluable technical support from the staffs of collaborating
institutions. This project was supported in part by Bundesministerium f\"ur Bildung, 
Wissenschaft, Forschung und Technologie, Consejo Nacional de 
Ciencia y Tecnolog\'{\i}a {\nobreak (CONACyT)},
Conselho Nacional de Desenvolvimento Cient\'{\i}fico e Tecnol\'ogico,
Fondo de Apoyo a la Investigaci\'on (UASLP),
Funda\c{c}\~ao de Amparo \`a Pesquisa do Estado de S\~ao Paulo (FAPESP),
the Israel Science Foundation founded by the Israel Academy of Sciences and 
Humanities, Istituto Nazionale di Fisica Nucleare (INFN),
the International Science Foundation (ISF),
the National Science Foundation, NATO,
the Russian Academy of Science, the Russian Ministry of Science and Technology,
the Russian Foundation for Basic Research (research project No.~11-02-01302-a),  
the Secretar\'{\i}a de Educaci\'on P\'ublica (Mexico),
the Turkish Scientific and Technological Research Board (T\"{U}B\.ITAK),
and the U.S.\ Department of Energy. We thank ITEP and National Research 
Nuclear University MEPhI (Moscow Engineering Physics Institute) for providing 
computing powers and support for data analysis and simulations.

The authors also would like to thank Prof. Michael Lisa and
Prof. Richard Lednick{\'y} for helpful comments and fruitful
discussions.





\bibliographystyle{elsarticle-num}
\bibliography{kaonKtDep}



\end{document}

%% file: selex_authors.tex

\address[bu]{Bogazici University, Bebek 80815 Istanbul, Turkey}
\address[cmu]{Carnegie-Mellon University, Pittsburgh, PA 15213, U.S.A.}
\address[cbpf]{Centro Brasileiro de Pesquisas F\'{\i}sicas, Rio de Janeiro, Brazil}
\address[fnal]{Fermi National Accelerator Laboratory, Batavia, IL 60510, U.S.A.}
\address[ihep]{Institute for High Energy Physics, Protvino, Russia}
\address[itep]{Institute of Theoretical and Experimental Physics, Moscow, Russia}
\address[mpik]{Max-Planck-Institut f\"ur Kernphysik, 69117 Heidelberg, Germany}
\address[msu]{Moscow State University, Moscow, Russia}
\address[mephi]{National Research Nuclear University MEPhI (Moscow Engineering Physics Institute), Moscow, Russia}
\address[pnpi]{Petersburg Nuclear Physics Institute, St.\ Petersburg, Russia}
\address[tau]{Tel Aviv University, 69978 Ramat Aviv, Israel}
\address[uaslp]{Universidad Aut\'onoma de San Luis Potos\'{\i}, San Luis Potos\'{\i}, Mexico}
\address[ub]{University of Bristol, Bristol BS8~1TL, United Kingdom}
\address[ui]{University of Iowa, Iowa City, IA 52242, U.S.A.}
\address[umf]{University of Michigan-Flint, Flint, MI 48502, U.S.A.}
\address[ur]{University of Rome ``La Sapienza'' and INFN, Rome, Italy}
\address[usp]{University of S\~ao Paulo, S\~ao Paulo, Brazil}
\address[ut]{University of Trieste and INFN, Trieste, Italy}

\fntext[cross]{Deceased}
\fntext[uci]{Present address: University of California at Irvine, Irvine, CA 92697, USA}
\fntext[amtc]{Present address: Advanced Mask Technology Center, Dresden, Germany}
\fntext[ifuec]{Present address: Instituto de F\'{\i}sica da Universidade Estadual de Campinas, UNICAMP, SP, Brazil}
\fntext[ku]{Present address: Kafkas University, Kars, Turkey}
\fntext[pdtum]{Present address: Physik-Department, Technische Universit\"at M\"unchen, 85748 Garching, Germany}
\fntext[bcg]{Present address: The Boston Consulting Group, M\"unchen, Germany}
\fntext[lt]{Present address: Lucent Technologies, Naperville, IL}
\fntext[bh]{Present address: Baxter Healthcare, Round Lake IL}
\fntext[nrcn]{Present address: NRCN, 84190 Beer-Sheva, Israel}
\fntext[sdu]{Present address: S\"uleyman Demirel Universitesi, Isparta, Turkey}
\fntext[doe]{Present address: DOE, Germantown, MD}
\fntext[solid]{Present address: Solidum, Ottawa, Ontario, Canada}
\fntext[sh]{Present address: Siemens Healthcare, Erlangen, Germany}
\fntext[duke]{Present address: Duke University, Durham, NC 27708, USA}
\fntext[ina]{Present address: Instituto Nacional de Astrof\'{\i}sica, \'Optica y Electr\'onica, Tonantzintla, Mexico}
\fntext[snolab]{Present address: SNOLAB}
\fntext[aigit]{Present address: Allianz Insurance Group IT, M\"unchen, Germany}


\author[mephi]{G.A.~Nigmatkulov}
\ead{nigmatkulov@gmail.com}
%
\author[mephi,cross]{A.K.~Ponosov}
%
\author[ui]{U.~Akgun}
%
\author[pnpi]{G.~Alkhazov}
%
\author[uaslp]{J.~Amaro-Reyes}
%
\author[itep]{A.~Asratyan}
%
\author[pnpi,cross]{A.G.~Atamantchouk}
%
\author[ui]{A.S.~Ayan}
%
\author[itep,cross]{M.Y.~Balatz}
%
\author[uaslp]{A.~Blanco-Covarrubias}
%
% ...
%
